\documentclass{PoS}

\title{Low Energy Background Spectrum in CDMSlite}

\ShortTitle{Low Energy Background Spectrum in CDMSlite}

    \author{\speaker{D. Barker} on behalf of the SuperCDMS Collaboration\thanks{\texttt{http://cdms.berkeley.edu/cdms\_collab.html}}\\
       School of Physics and Astronomy, University of Minnesota, Minneapolis, Minnesota 55455, USA\\
       E-mail: \email{barker@physics.umn.edu}}

\abstract{
One trend in dark matter direct detection is the development of techniques which will lower experimental thresholds and achieve sensitivity to light mass dark matter particles. In doing so, it is necessary to have an understanding of the low energy spectrum of the major background components. Geant4 has a number of specialized low energy physics processes that can be implemented when simulating an experimental geometry. To understand this low energy region for the Super Cryogenic Dark Matter Search (SuperCDMS), a variety of these models have been simulated and compared against theoretical calculations and SuperCDMS calibration data. Most of the low energy processes include a more complete description of the atomic structure, allowing us to observe the phenomenon of Compton steps in the simulation. An important application of this low energy background modeling is for the SuperCDMS low ionization threshold experiment (CDMSlite). CDMSlite has reached world-leading sensitivities in the search for low mass weakly interacting massive particle (WIMP) dark matter. Using Neganov-Trofinov-Luke phonon amplification, CDMSlite has achieved a threshold of less than 60 eV for electron recoils. The dark matter sensitivity of the CDMSlite-Soudan data can be improved by understanding and modeling the experimental backgrounds down to this threshold. Development of the machinery for creating a low energy background model will also be useful in the future SuperCDMS-SNOLAB experiment which will run multiple high voltage detectors in CDMSlite mode. 
}

\FullConference{38th International Conference on High Energy Physics\\
                  3-10 August 2016\\
                 Chicago, USA}

\begin{document}

\section{Introduction}
To extend the dark matter search to low mass WIMP particles, \nobreak{SuperCDMS} has run several detectors at a higher bias voltage in a mode called the low ionization threshold experiment (\nobreak{CDMSlite})~\cite{lite2}. This method achieves a lower threshold through the use of the Neganov-Trofimov-Luke (NTL) effect, which amplifies the phonon signal of charges drifting through an electric field in a crystal~\cite{ntl}. By using this technique, the discrimination between nuclear recoils and electron recoils is exchanged for a threshold of tens of eVs. Compared with the standard SuperCDMS iZIP technology, \nobreak{CDMSlite} is background-limited, with the electron recoil and surface events dominating the data~\cite{lite2}. Plans to utilize NTL amplification for the new \nobreak{SuperCDMS}-SNOLAB experiment will require extensive background modeling to perform likelihood fits.

\nobreak{SuperCDMS} uses the Geant4~\cite{geant4} simulation package to estimate the rate of events from background environmental and cosmogenic sources. The `Shielding' Physics List is widely used, and agrees with the data in the region of a few keV and above~\cite{reana}. As \nobreak{SuperCDMS} continues to reduce its energy detection threshold, it becomes increasingly important to understand the physics processes that dominate at low energy and verify that the simulation is giving expected results.

\section{Geant4 Low Energy Physics}
A low energy electromagnetic physics list (\texttt{G4EmStandardPhysics\_option4}) is available in \nobreak{Geant4} with a variety of models that have been specifically designed for the region below a few keV~\cite{geant4}. These models consider the atomic structure of the atom, atomic de-excitation, and properties of the bound electron. A model of particular interest is the Monash Compton model (\texttt{G4LowEPComptonModel})~\cite{Monash}. The Monash Compton model calculates properties of the outgoing gamma and recoiling electron by modifying standard assumptions: the initial electron is not at rest, and the final state products are not confined to a two dimensional plane. This more complete treatment is necessary to replicate the phenomenon of `Compton steps': step features created in the low energy spectrum because the detector will collect at least the binding energy of any freed electron. For example, the electrons in the \textit{K}-shell of germanium have a binding energy of 11.1\,keV. This energy is deposited in the detector due to the reorganization of the electron shells, along with any additional energy that is given to the freed electron by the incoming gamma. Thus, an electron from the \textit{K}-shell can never deposit less than 11.1\,keV in the detector, and likewise for electrons in the other atomic shells. Naively, we expect the number of electrons in each shell to determine the relative size of the steps, however details of the electron wave functions can also affect the size. The Compton steps have been observed previously in silicon detectors~\cite{damic}.

\begin{figure}[t!]
	\centering
		\includegraphics[width=0.38\textwidth]{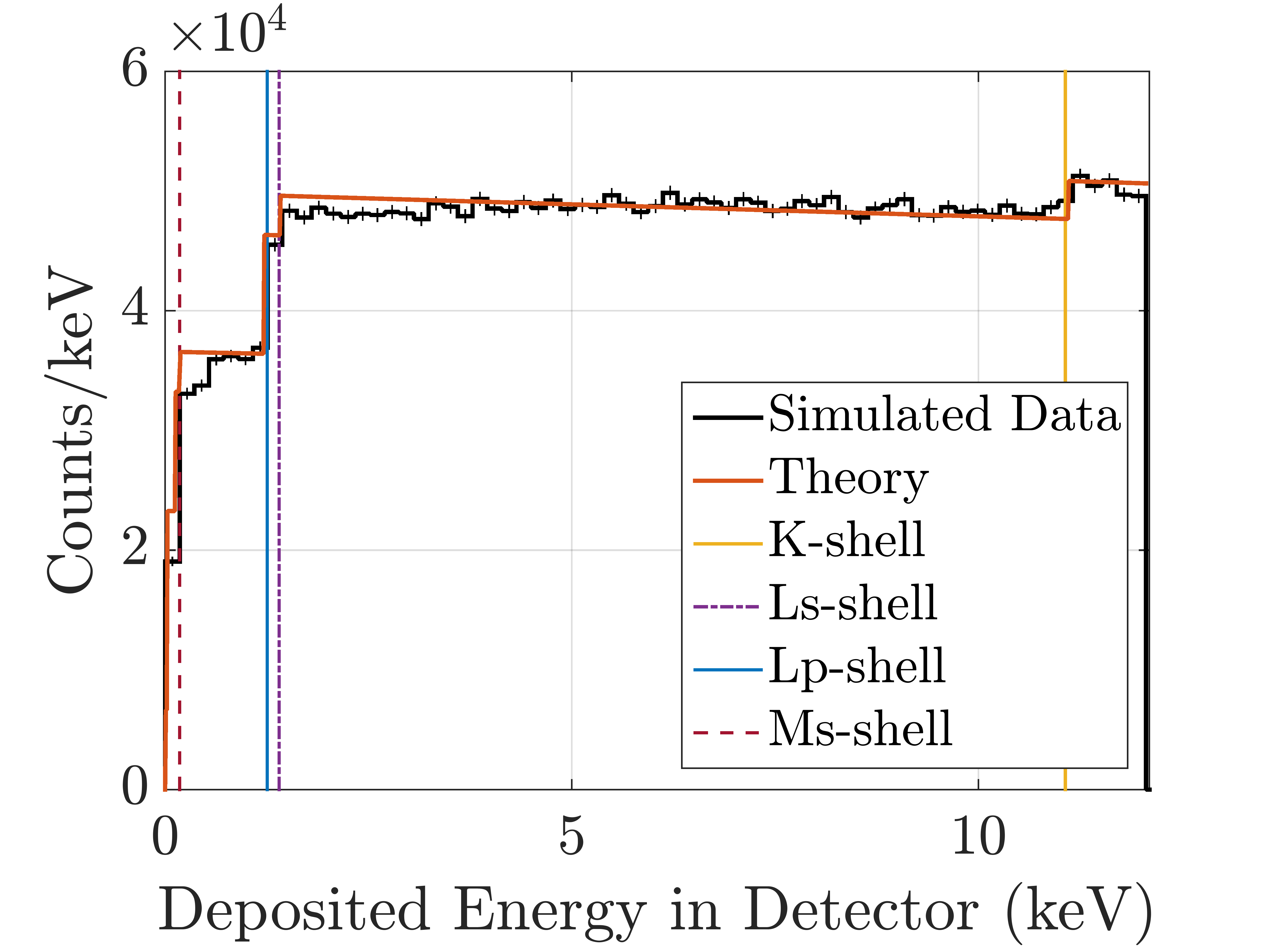}
		\includegraphics[width=0.38\textwidth]{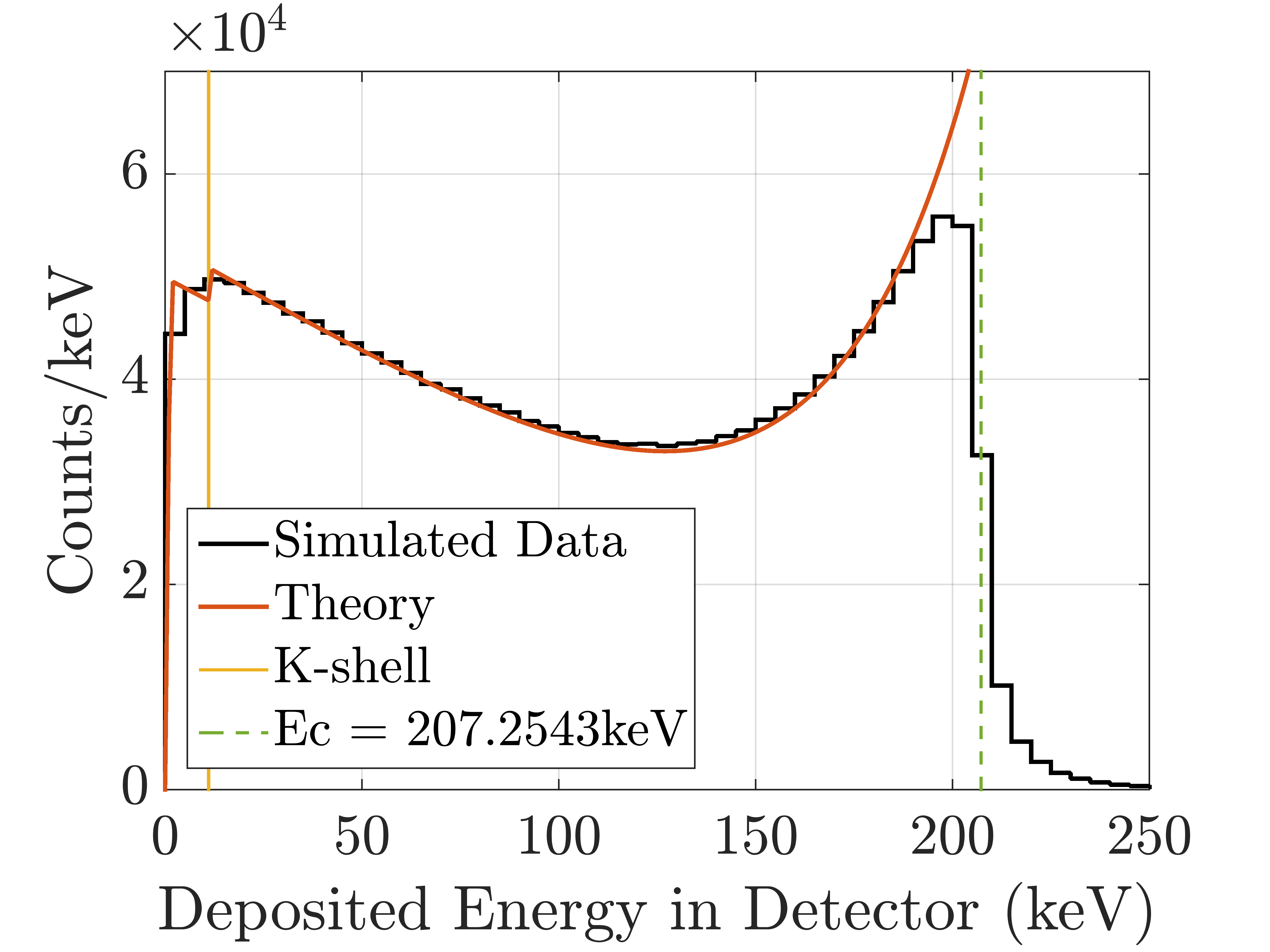}
	\caption{Simulated data from incoming gammas of 356\,keV on a germanium target compared to a modified Klein-Nishina theoretical model. \textit{Left:} The low energy region demonstrating the structure from the Compton steps. \textit{Right:} The entire range of energy deposition where the tail above the Compton edge is visible.}
	\label{fig:Monash}
\end{figure}

With this more detailed physics list, a simulation was completed with 356\,keV gammas, the most prominent energy of gammas from $^{133}$Ba decay and the primary electron recoil source for \nobreak{SuperCDMS}-Soudan. These are incident isotropically on a germanium detector the size of a \nobreak{SuperCDMS} iZIP. The effect of the Monash Compton model can be seen in Figure~\ref{fig:Monash}. The simulation spectrum is compared to the Klein-Nishina equation~\cite{kn}, modified using the naive assumption above, to emulate the Compton steps. The left plot shows the low energy steps that appear at the binding energies of germanium. The theoretical description approximately matches the simulation in this energy range; the discrepancies are mainly due to the negative slope of the Klein-Nishina, which is not reproduced in simulation, and the naive step size. On the right hand side of Figure~\ref{fig:Monash}, a high energy tail is visible above the Compton cutoff. If the momentum direction of the bound electron and the direction of the incoming gamma are aligned, it is possible for the electron to have more kinetic energy than is predicted by the calculation with the electron at rest. The theoretical calculation matches the full energy range to about 200\,keV, beyond which events are smeared due to the momentum of the bound electron. 

\section{Comparison with SuperCDMS Data}
A set of $^{133}$Ba high statistics data was collected over the course of about a month. From this data, the region below 20 keV was considered for each detector. Quality cuts were applied to remove poorly reconstructed events. Only events that triggered a detector are used. Since there is no applied cut to define a fiducial volume for the detectors, there is some incomplete signal collection and a resulting effect on the spectra. The data from all detectors was combined and compared to a Geant4.10 simulation of the \nobreak{SuperCDMS}-Soudan experiment during $^{133}$Ba calibration. The energy deposited in the detectors in the simulation were considered individually in order to apply a detector dependent energy resolution and then combined. 

\begin{figure}[t]
	\centering
		\includegraphics[width=0.32\textwidth]{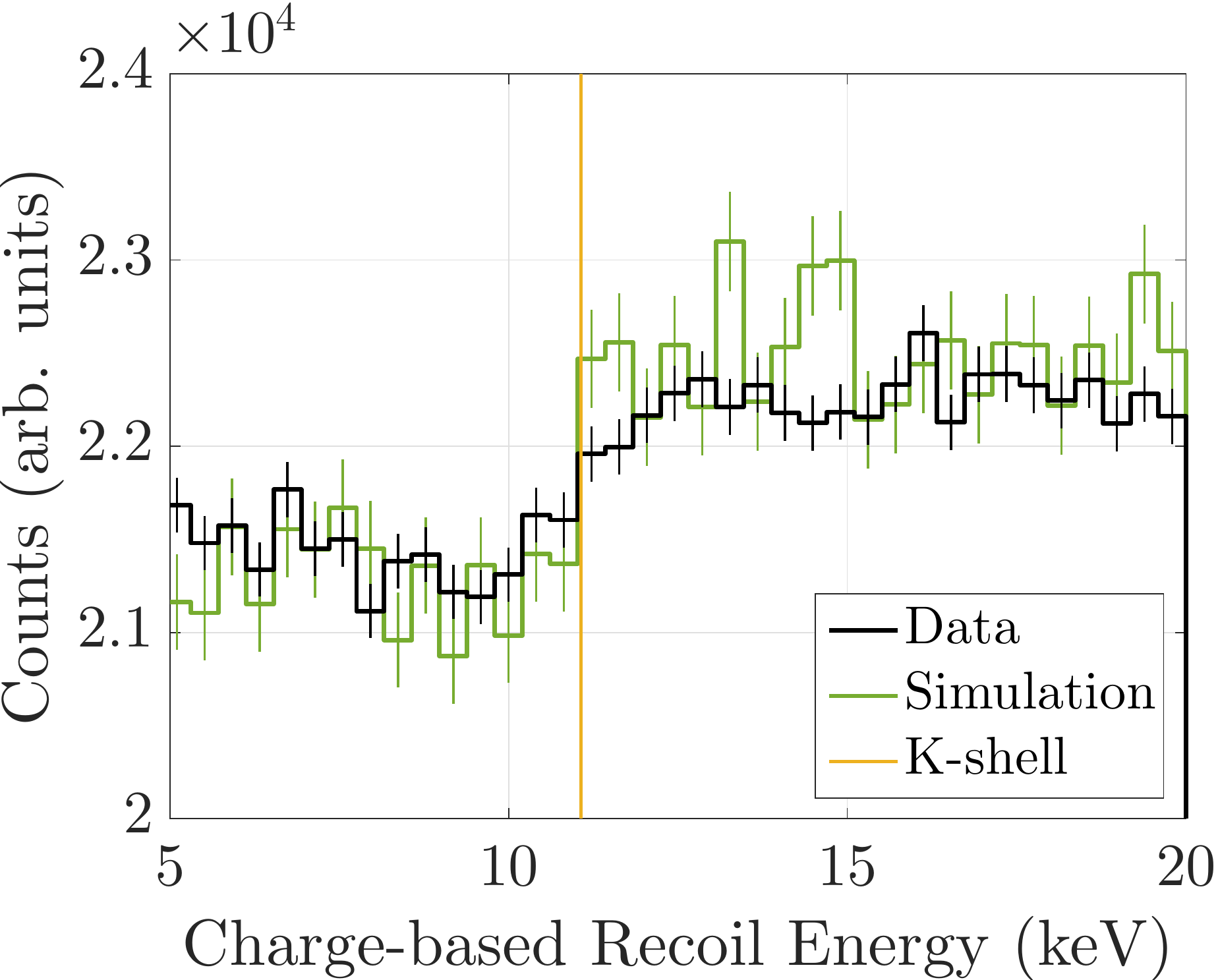}
		\hspace{0.1cm}
		\includegraphics[width=0.32\textwidth]{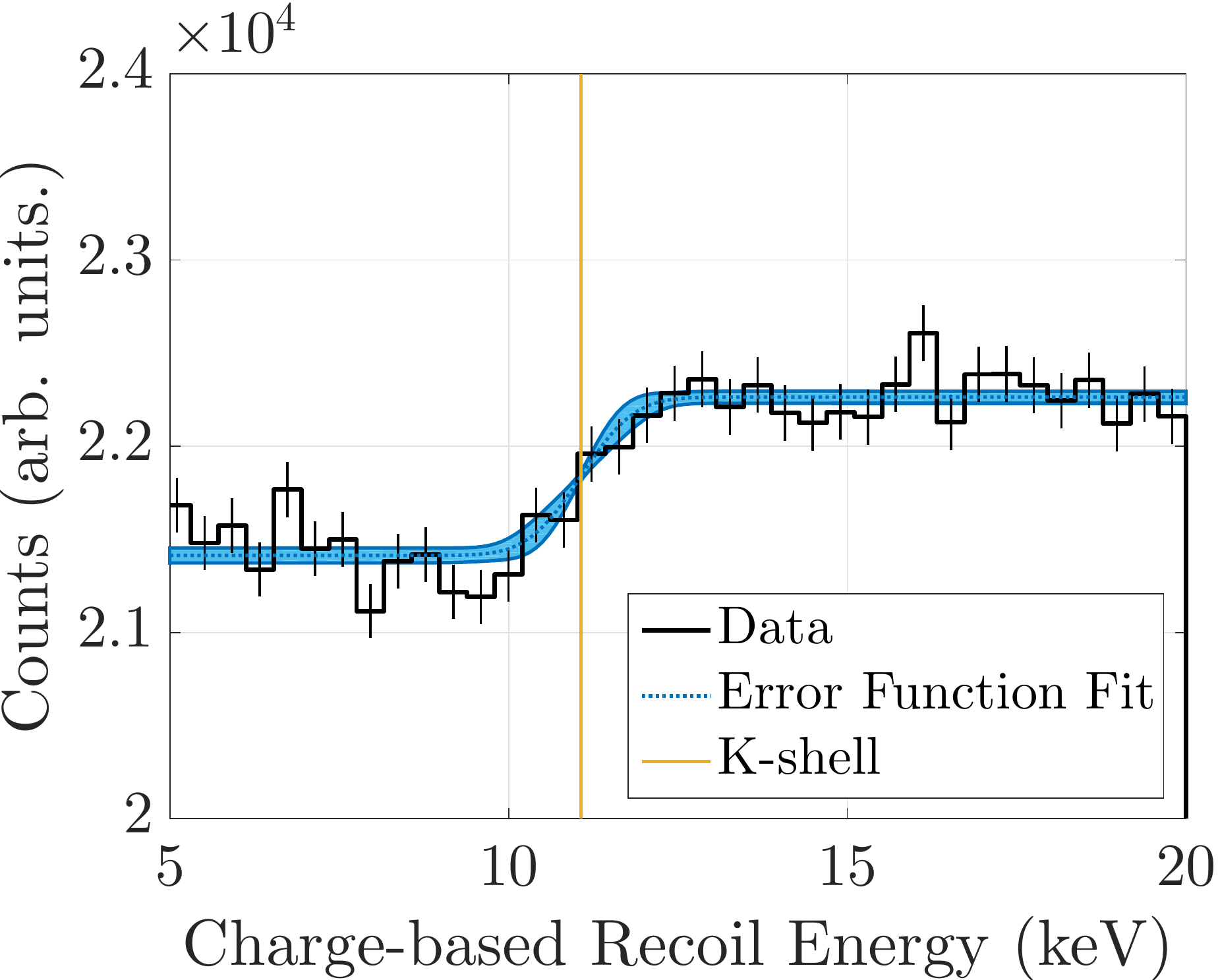}
		\hspace{0.1cm}
		\includegraphics[width=0.32\textwidth]{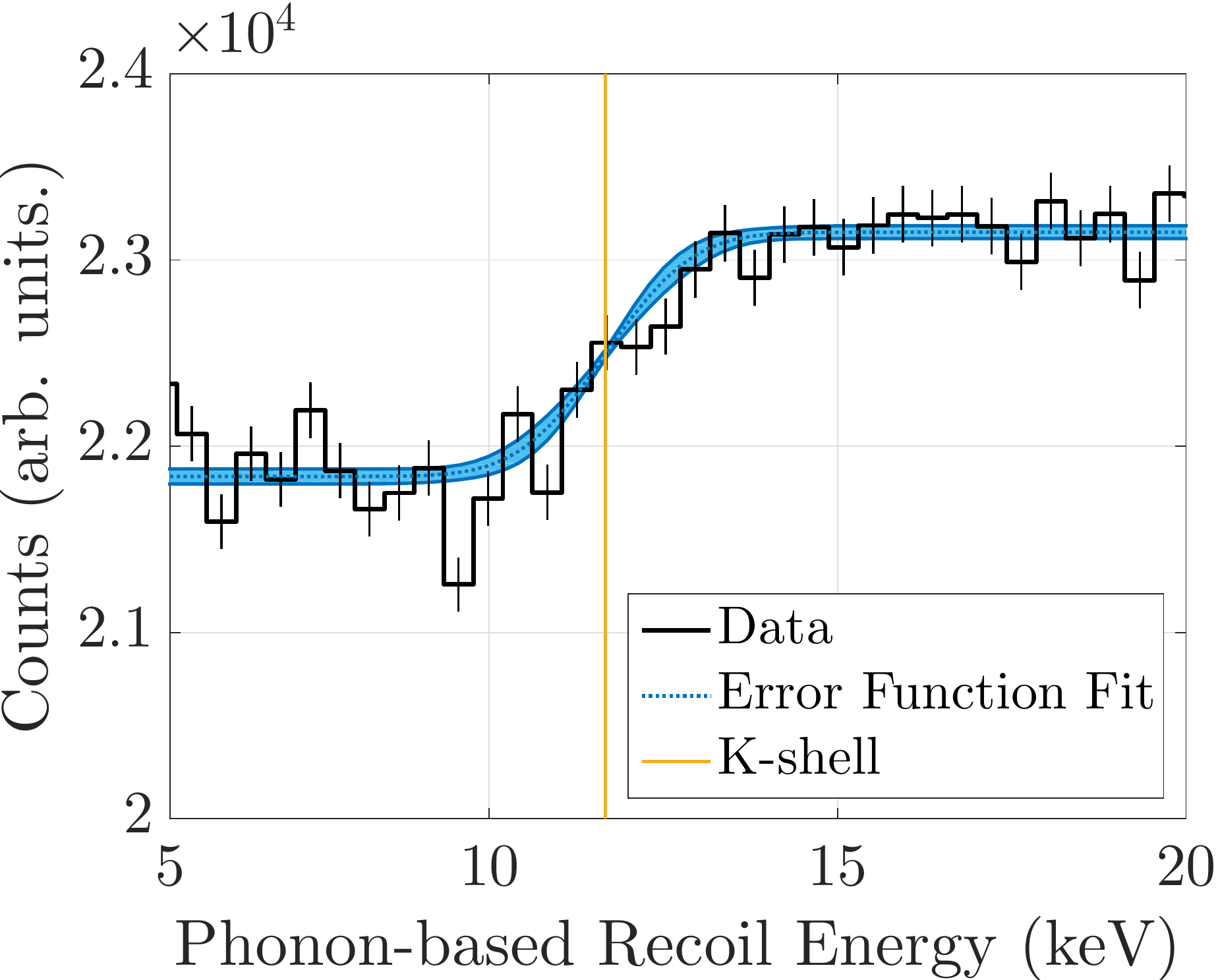}
	\caption{\textit{Left:} The low energy region of $^{133}$Ba calibration data (black) with simulated data from Geant4 using the \texttt{G4LowEPComptonModel} (green). \textit{Center:} Error function fit to the calibration data with the charge energy estimator. \textit{Right:} Error function fit to the calibration data with the phonon energy estimator. The spectrum is not analyzed below 5 keV where triggers on noise and uncertainties in cut acceptances bias measurements of the Compton steps.}
	\label{fig:data}
\end{figure}

Figure~\ref{fig:data} \textit{Left} shows the comparison between simulation and data. The \textit{K}-shell of germanium, at 11.1\,keV, is indicated by the vertical line. There is a flat plateau above 11.1\,keV and a second lower one below, where the observed deviations on the plateaus are consistent with statistical fluctuations of a flat distribution. Good agreement is observed between the data and simulation. The data is fit with an error function and constant offset. This can be seen in Figure~\ref{fig:data} \textit{Center} and \textit{Right} for the charge and phonon energy estimators respectively. The amplitude of the error function is directly related to the \textit{K}-shell step size. Considering only the number of available electrons in the \textit{K}-shell, the expected relative step size is naively 6.25\%. From the simulation, the expected relative size is smaller at 5.3 $\pm$ 0.3\% due to the consideration of the electron properties. The measured difference in rates in data are 3.7 $\pm$ 0.2\% in charge energy and 5.4 $\pm$ 0.2\% in phonon energy. All given errors are statistical. The systematic uncertainty from varying the slope of the error function fit are on the same order. However, other systematics, like incomplete signal collection, have yet to be considered. Figure~\ref{fig:data} gives additional experimental evidence of the low energy Compton steps. In addition, the agreement with simulation gives confidence to the use of \texttt{G4LowEPComptonModel} and \texttt{G4EmStandardPhysics\_option4} for future \nobreak{SuperCDMS} studies.

\section{Future Work}
Continued investigation of the Compton steps is ongoing, along with the development of a generic analytic expression for the Compton background for use with \nobreak{CDMSlite} data. However, there are other backgrounds that need to be considered. The activation peaks due to calibration with $^{252}$Cf are well understood~\cite{lite2}. These, as well as the cosmogenic activation due to the exposure of the crystal during fabrication above ground, have known distributions that can be incorporated into a background model. Finally, the surface backgrounds from radon or other contaminants on the top and bottom faces of the detectors need to be considered at low energies. All these sources must be included in order to implement a background model for the future \nobreak{CDMSlite} analysis. This work is also important as the first step to developing the tools to be used for understanding the backgrounds of the future \nobreak{SuperCDMS}-SNOLAB experiment.

\end{document}